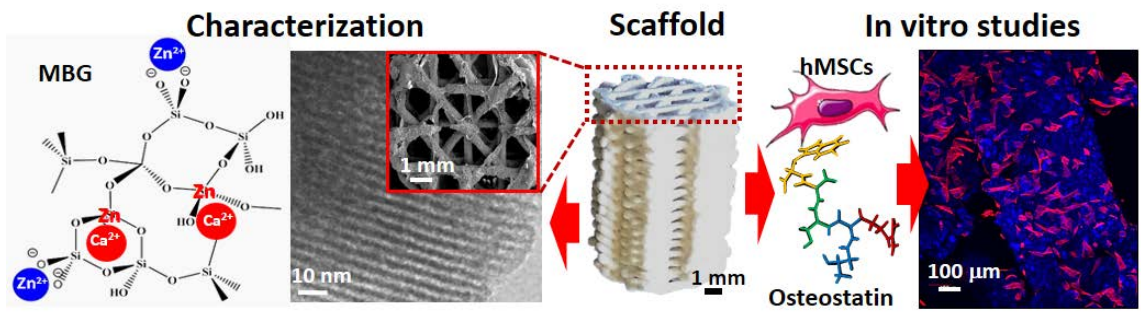

**Graphical abstract**

# Osteostatin potentiates the bioactivity of mesoporous glass scaffolds containing $Zn^{2+}$ ions in human mesenchymal stem cell cultures

Heras C.[a], Sanchez-Salcedo S.[a,b,*], Lozano D.[a,b], Peña J.[a], Esbrit P.[a], Vallet-Regi M.[a,b], Salinas A. J.[a,b,*]

[a] *Departamento de Química en Ciencias Farmacéuticas, Facultad de Farmacia, Universidad Complutense de Madrid, UCM, Instituto de Investigación Hospital 12 de Octubre, i+12, 28040 Madrid, Spain.*
[b] *Networking Research Center on Bioengineering, Biomaterials and Nanomedicine (CIBER-BBN), Spain.*
salinas@ucm.es

**Abstract**

There is an urgent need of biosynthetic bone grafts with enhanced osteogenic capacity. In this study, we describe the design of hierarchical meso-macroporous 3D-scaffolds based on mesoporous bioactive glasses (MBGs), enriched with the peptide osteostatin and $Zn^{2+}$ ions, and their osteogenic effect on human mesenchymal stem cells (hMSCs) as a preclinical strategy in bone regeneration. The MBG compositions investigated were 80%$SiO_2$–15%CaO–5%$P_2O_5$ (in mol-%) Blank (BL), and two analogous glasses containing 4% ZnO (4ZN) and 5% ZnO (5ZN). By using additive fabrication techniques, scaffolds exhibiting hierarchical porosity: mesopores (around 4 nm), macropores (1-600 μm) and big channels (∼ 1000 μm), were prepared. These MBG scaffolds with or without osteostatin were evaluated in cell cultures of hMSCs. Zinc promoted hMSCs colonization (both the surface and inside) of MBG scaffolds. Moreover, $Zn^{2+}$ ions and osteostatin together, but not independently, in the scaffolds were found to induce the osteoblast differentiation genes runt related transcription factor-2 (RUNX2) and alkaline phosphatase (ALP) in hMSCs after 7 d of culture in the absence of an osteogenic differentiation-promoting medium. These results add credence to the combined use of zinc and osteostatin as an effective strategy for bone regeneration applications.

**Statement of significance**

Mesoporous bioactive glasses (MBGs) are bioceramics whose unique properties make them excellent materials for bone tissue engineering. Physico-chemical characterization of MBGs as scaffolds made by rapid prototyping, then doped with zinc (potential osteogenic, angiogenic and bactericidal ion) and loaded with osteostatin (osteogenic peptide) are described. These Zn-MBGs scaffolds showed 3D hierarchical meso-macroporous structure that enables to host and release osteostatin. When decorated with human mesenchymal stem cells (hMSCs), MBGs scaffolds enriched with both zinc and osteostatin exhibited a synergetic effect to enhance hMSC growth, and also their osteogenic differentiation without addition of classical osteoblastic differentiation factors to the culture medium. This novel strategy has a great potential for use in bone tissue engineering.

# 1. Introduction

Bone tissue regenerates spontaneously after damage. However, in certain physiological situations or upon extensive damage, normal regeneration mechanisms are not enough to regenerate bone. In those cases, bone regeneration needs to be enhanced by different maneuvers such as tissue engineering approaches [1,2]. Recently, $SiO_2$–$CaO$–$P_2O_5$ mesoporous bioactive glasses (MBGs) gained much interest for their excellent properties: bone regenerative capacity; high ordered mesoporous structure enabling them to host and release bone promoting agents [3,4]; and huge surface area and pore volume which provide fast in vitro responses [5,6]. The regenerative potential of bioactive glasses depends in part on their ionic dissolution products (Si, Ca, P) [7]. In addition, the bioactive properties of MBGs can be improved by including additional metal ions such as $Cu^{2+}$, $Mg^{2+}$, $Ce^{3+}$, $Co^{2+}$, $Ga^{3+}$, or $Zn^{2+}$ into their structure. This strategy has attracted much attention by its simplicity, good stability and clinical safety for the proper range for each metallic ion. It has been demonstrated, both in vitro and in vivo, that dissolution of these metal-enriched bioactive glasses induced an increased intracellular ion concentration and subsequent activation of intracellular signaling pathways; this leads to an increased gene expression in osteoprogenitor cells which promotes bone regeneration [8,9]. Zinc ions appear to be a good option for that purpose due to their capacity to increase osteogenesis [1,10] angiogenesis [11], and antimicrobial properties [12,13].

The high surface area and pore volume of MBGs are suitable to host biologically active molecules that, combined with their interconnected macropores and channels (100–1000 μm), allows tissue ingrowth and vascularization [14]. In this context, rapid prototyping processing has been shown to be a successful technique of stereolithography for obtaining meso-macroporous bioactive scaffolds. Moreover, coating these scaffold systems with biodegradable and biocompatible biopolymers improves their handling and in vivo performance in terms of adsorption or release of ions in the surrounding medium while preserving their morphology and integrity [15]. Gelatin is a noncytotoxic FDA-approved biopolymer that can be formulated as a hydrogel with high swelling ability by crosslinking with glutaraldehyde (GA) [15]. This formulation appears to be an ideal bone filling agent for its flexibility and capacity to exert pressure in contact with body fluids, and easy surgical handling [16].

The bioactivity of these scaffolds could also be further improved by loading different osteogenic molecules. In this regard, parathyroid hormone (PTH)-related protein (PTHrP) is an emerging bone regenerating agent whose N-terminal region is homologous to PTH, whereas its PTH-unrelated C-terminal region includes a highly conserved 107-111 sequence (named osteostatin). When intermittently administered, N-terminal PTHrP analogues have been shown to induce bone anabolism in rodents and humans [17,18]. On the other hand, C-terminal PTHrP peptides exhibit antiresorptive properties [19], but they also have osteogenic activity in different in vitro and in vivo scenarios [20,23]. Very recently, our group reported a straightforward method to synthesize MBG disks enriched with zinc and osteostatin, exhibiting an improved capacity to stimulate osteoblastic cell growth and function [24].

In the present study, we aimed to explore the hypothesis that the combination of zinc-enriched MBGs and osteostatin is an interesting approach for bone tissue engineering. Thus, hierarchical meso-macroporous MBG scaffolds were synthesized based on 80% $SiO_2$–15%CaO–5%$P_2O_5$ (mol-%) system: Blank as a control (BL), and including 4% or 5% of ZnO (4ZN and 5ZN), coated with GA crosslinked gelatin, and loaded (or not) with osteostatin. We used human mesenchymal stem cells (hMSCs) to decorate these scaffolds. The putative advantage of including zinc and osteostatin together in these MBG scaffolds to enhance cell growth and osteogenic differentiation of hMSCs was evaluated.

## 2. Materials and methods

MBGs were synthesized by the evaporation induced self-assembly (EISA) technique, and MBG scaffolds as 3D structures were obtained by rapid prototyping techniques. They were characterized physico-chemically, and their swelling capacity and ion release time course were also assayed. These scaffolds were thereafter seeded with human mesenchymal stem cells followed by evaluation of cell toxicity and cell proliferation as well as osteogenic differentiation as described below.

*2.1 Synthesis of bioactive mesoporous glasses (MBGs)*

Mesoporous glasses ($SiO_2$–CaO–$P_2O_5$) were synthesized using the EISA (evaporation induced self-assembly) method. $SiO_2$, $P_2O_5$ and CaO sources came from tetraethylorthosilicate (TEOS), triethyl phosphate (TEP) and calcium nitrate

($Ca(NO_3)_2 \cdot 4H_2O$) respectively, and Pluronic® P123 was used as a structure directing agent. All of them purchased from Sigma-Aldrich (St. Louis, USA). The reaction was initiated with a mixture of 4.5 g of Pluronic® P123 dissolved in 67.5 mL of absolute ethanol (EtOH) (Panreac Química SLU; Castellar del Vallés, Barcelona, Spain) and 1.12 mL of 0.5N $HNO_3$. One h later, 8.90 mL of TEOS were added, and 0.71 mL of TEP and 1.10 g of $Ca(NO_3)_2 \cdot 4H_2O$ were subsequently incorporated every 3 h thereafter. The reaction was carried out overnight, but 0.60 g (4ZN) or 0.75 g (5ZN) of $Zn(NO_3)_2 \cdot 6H_2O$ was added after 1 h of reaction to obtain zinc doped scaffolds (Table 1). Twenty four h later, the resulting sols were placed on 9-cm Petri dishes to continue with the EISA method. The gelation process was carried out for 4 d at 30 ºC, and the resulting homogeneous and transparent membranes were calcined at 700 ºC for 6 h to obtain the glasses.

*2.2 Fabrication of MBG scaffolds by rapid prototyping*

The so-obtained glasses were gently grinded in a mortar and sieved through a 40 µm mesh to get the glass powders. Then, 4.5 g of MBG powder was dispersed in 37 mL of dichloromethane (DCM), followed by addition of 2.5 g of ε-polycaprolactone (PCL) (Mw = 58,000 Da) (Sigma-Aldrich) in 25 mL of DCM. The mixture was stirred at room temperature until cementing with the right consistency for handling in a 3D printer rapid prototyping equipment 3D Bioplotter™ (EnvisionTEC, Gladbeck, Germany) printer. Thus, the paste was placed in a polyethylene cartridge with a 0.55-mm dispensing tip, setting the rod spacing to 1.5 mm. Each consecutive layer was correspondingly rotated to a 45º angle to generate two types of rhombohedral channels. The tip speed was set at 300 or 50 mm/min in the horizontal or the vertical plane, respectively. The dimensions of the obtained cylindrical scaffolds were 6 mm diameter x 10 mm height. The scaffolds were dried at 37 ºC for 24 h to evaporate DCM and then treated at 700 ºC for 6 h to remove PCL. For some characterization techniques, such as nitrogen adsorption and mercury porosimetry, smaller scaffolds (6 mm diameter x 5 mm height) were generated. The resulting scaffolds were coated with glutaraldehyde (GA) crosslinked gelatin (2.4 wt-%) to improve their mechanical properties (BL-GE, 4ZN-GE and 5ZN-GE scaffolds). Thermogravimetric analysis (TGA) was carried out in a Perkin-Elmer Pyris Diamond TG/DTA instrument (Waltham, MA, USA) to determine any residual content of PCL in the scaffolds. TGA was performed between 30 and 900 °C in air at a flow rate of 100 mL/min and a heating rate of 10 °C/min. Some of these scaffolds were loaded with osteostatin as described in section 2.6. (BL-GE+OST, 4ZN-GE+OST and 5ZN-GE+OST scaffolds).

*2.3 Meso-macroporous 3D scaffolds characterization*

Nitrogen adsorption measurements were carried out in the calcined, GA crosslinked gelatin coated and loaded with osteostatin scaffolds to quantify pore diameter, pore volume and surface area. Analyses were performed at –196 °C using a Micromeritics ASAP 2020 (Micromeritics, Norcross, USA). First, 100–150 mg of each scaffold was degassed for 24 h at 200 °C under vacuum (< 0.3 kPa). The surface area ($S_{BET}$) was determined by the Brunauer–Emmett–Teller (BET) technique. The total pore volume ($V_T$) was calculated from the amount of $N_2$ adsorbed at a relative pressure of 0.97. The possible existence of micropores (< 2 nm) was analyzed by the *t*-plot method. The average mesopore diameter was determined from the adsorption branch of the isotherm by the Barrett–Joyner–Halenda (BJH) method. The presence of an ordered mesoporous arrangement was investigated by small angle X-ray diffraction (SA-XRD) in a Philips X'Pert diffractometer (Eindhoven, The Netherlands) equipped with a Cu $K_\alpha$ radiation source (wavelength 1.5418 Å). SA-XRD patterns were registered in the $2\theta$ range between 0.6° and 8°, with a step size of 0.02° and a counting time of 0.5 s *per* step. Mesoporous channel arrangement was also evaluated by transmission electron microscopy (TEM) of crushed scaffolds carried out in a JEOL2100 electron microscope operating at 200 kV (Tokyo, Japan). Fourier transform infrared FTIR spectroscopy was used to assess the composition of the coated and not coated scaffolds with GA crosslinked gelatin. The latter measurements were performed in a Thermo Scientific Nicolet iS10 apparatus (Waltham, MA, USA) equipped with a SMART Golden Gate® attenuated total reflection ATR, a diffuse reflectance accessory.

To study the interconnected macroporosity of MBG scaffolds and its chemical composition, a JEOL JSM-6400 scanning electron microscope (SEM) (Tokyo, Japan) operating at 20 kV and equipped with an Oxford Instruments INCA Energy Dispersive X-ray (EDX) spectrometer was used. SEM micrographs from the intact scaffold surface and from a scaffold fracture were obtained to describe the morphology of the calcined scaffolds and also of those coated with GA crosslinked with gelatin. The organic matter content was quantified by TGA. In addition, smaller scaffolds (6 mm x 5 mm) with a weight of 100–150 mg were studied by mercury intrusion porosimetry as complementary information about their pore diameter and surface area. This analysis was done with a Micromeritics Autopore IV 9500 device (Norcross, GA, USA).

Nuclear magnetic resonance (NMR) was performed to study the atomic environments of silicon and phosphorus atoms in the tested materials. A Bruker Avance AW-400WB spectrometer (Karlsruhe, Germany) was used. Samples were spun at 10 and 6 kHz for $^{29}$Si and $^{31}$P setting the spectrometer frequencies to 79.5 and 162.0 MHz. Chemical shift values were referenced to tetramethylsilane (TMS) and $H_3PO_4$ for $^{29}$Si and $^{31}$P.

*2.4 Swelling assay*

The swelling behavior of BL-GE, 4ZN-GE and 5ZN-GE scaffolds in distilled water at 37 ºC was determined by gravimetric analyses (%$W_{REHYDRATION}$) as follows:

%$W_{REHYDRATION} = 100 \times (W_t - W_d / W_d)$  (Eq. 1)

Where Wt is the weight of the rehydrated sample at time t, and Wd, the weight of the dried sample. Wt values were determined at different times until the scaffolds reached the swelling ratio equilibrium.

*2.5 Ion release assay*

To measure the solubility degree of the scaffolds, they were placed on 12-well transwells plates in contact with Dulbecco's Modified Eagle's medium with 10% fetal bovine serum and 1% penicillin-streptomycin, under shaking conditions for 10 d. The medium was extracted every day to measure the ion content (Ca, P and Zn) by inductively coupled plasma/optical emission spectrometry (ICP/OES) using an OPTIMA 3300 DV Perkin Elmer device (Waltham, MA, USA). The concentration of each ion was determined in two different samples measured by triplicate.

*2.6 In vitro tests*

Previously to the *in vitro* studies below, the scaffolds were sterilized using UV light for 60 min with periodical rotation of the sample.

Human MSCs (hMSCs) were cultured in mesenchymal stem cell basal medium (MSCBM; Lonza, Walkersville, MD, USA). Cells were washed with phosphate buffered saline (PBS, pH 7.4) and then tripsinized with 5 mL of trypsin/0.25 % EDTA (Lonza). Cells were then centrifuged at 1200g for 7 min, and the resulting pellet was suspended in culture medium and added (100 µL) drop by drop onto each scaffold ($15 \cdot 10^4$ cells/scaffold) in a 24-well plate. Thereafter, 1.2 mL of MSCBM were added *per* well, and scaffolds were incubated at 37 ºC in 5% $CO_2$ for different times. Some of the scaffolds were previously loaded with osteostatin (Bachem, Bubendorf, Switzerland) by soaking in 2.5 mL of 100 nM peptide in PBS and continuous stirring for 24 h at 4 ºC. Peptide uptake and release were determined by measuring absorbance at 280 nm of the saline solution at different times.

*2.6.1 Morphological studies by confocal laser scanning microscopy*

Human MSCs attachment to the different tested scaffolds was studied by fluorescence microscopy. Each scaffold was rinsed twice in PBS and then fixed with 4% p-formaldehyde in PBS at 37 ºC for 20 min. Then, the scaffolds were washed again in PBS, and permeabilized in 0.5% Triton X-100 at 4 ºC for 5 min. Thereafter, non-specific binding was blocked with 1% bovine serum albumin (BSA) in PBS for 20 min at 37ºC. The samples were then incubated at 37ºC for 20 min with Atto 565-conjugated phalloidin (at 1:40 dilution; Molecular Probes, Thermo Fisher, Waltham, MA, USA), which stains actin filaments of the cell microskeleton. After washing with PBS, cell nuclei were stained with 3 µM 4´-6-diamino-2´-phenylindole dihydrochloride (DAPI) in PBS (Molecular Probes). Fluorescence microscopy was analyzed with a confocal laser scanning microscope OLYMPUS FV1200 (OLYMPUS, Tokyo, Japan), using a 60X FLUOR water dipping lens (NA=1.0). Micrographs were obtained using software 3D Imaris to project a single 2D image -converted into a TIF file- from the multiple Z sections by using an algorithm displaying the maximum pixel value of each Z 1-µm slice. DAPI and Atto 565-phalloidin staining were shown in blue and red respectively, whereas the scaffold reflection was shown in green.

*2.6.2 Cell growth*

Human MSC growth was determined after 5 and 7 d of cell culture by the Alamar Blue method (AbD Serotec, Oxford, UK), following manufacturer's instructions. Cells grown onto the scaffolds, before (BL-GE, 4ZN-GE, 5ZN-GE) and after osteostatin load (BL-GE+OST, 4ZN-GE+OST, 5ZN-GE+OST), were exposed to 10% Alamar Blue solution for 4 h. Then, fluorescence intensity was measured with excitation and emission wavelengths of 560 and 590 nm, respectively, by a BioTek Synergy 4 spectrometer (Winooski, VT, USA).

*2.6.3 Cell cytotoxicity*

Measurement of LDH activity released from hMSCs cells onto the scaffolds was used as a cytotoxicity assay. It is based on the reduction of nicotinamide adenine dinucleotide (NAD) by LDH [25], and the resulting colored compound is measured spectrophotometrically at 340 nm every 60 s for 4 min. This assay was carried out at day 10 of cell seeding by using a commercial kit (Spinreact, La Vall D'en Bas, Spain).

*2.6.4 Real Time PCR*

Total RNA was isolated from hMSCs seeded onto the scaffolds after 7 d of cell culture (Trizol, Invitrogen, Groningen, The Netherlands). Gene expression of two osteoblastic differentiation markers, alkaline phosphatase (ALP) and runt related transcription factor-2 (RUNX2) were analyzed by real time PCR using QuantStudio5 equipment and a described protocol [24] (Applied Biosystems-Thermo Scientific, Foster City, CA, USA). TaqMan$^{MGB}$ probes were obtained by Assay-by-Design$^{SM}$ (Applied Biosystems). mRNA copy numbers were calculated for each sample by using the cycle threshold (Ct) value, and glyceraldehyde-3-phosphate dehydrogenase (GAPDH) rRNA (a housekeeping gene) was amplified in parallel with the tested genes. The number of amplifications steps required to reach an arbitrary Ct was computed. The relative gene expression was represented by $2^{-\Delta\Delta Ct}$, where $\Delta\Delta Ct = \Delta Ct_{target\ gene} - \Delta Ct_{GAPDH}$. The fold change for the treatment was defined as the relative expression compared with control GAPDH expression, calculated as $2^{-\Delta\Delta Ct}$, where $\Delta\Delta Ct = \Delta C_{treatment} - \Delta C_{control}$.

*2.7. Statistical Analysis*

Results are expressed as mean ± SEM (standard error of the mean). Statistical analysis was performed with nonparametric Kruskal-Wallis test and post-hoc Dunn´s test. A value of $p < 0.05$ was considered significant.

**3. Results and discussion**

*3.1. Synthesis and shaping of 3D MBG scaffolds by rapid prototyping technique*

Mesoporous materials of the ternary $SiO_2$–$CaO$–$P_2O_5$ system were easily prepared in alcoholic medium under acidic conditions by using a non-ionic surfactant Pluronic® P123 as structure-directing agent and the EISA method, as described above.

TEOS, TEP, calcium nitrate tetrahydrate and zinc nitrate hexahydrate were used as $SiO_2$, $P_2O_5$, CaO and ZnO sources, respectively (Table 1). As described in detail in section 2.2, MBG scaffolds were obtained by rapid prototyping, using a previous computer-aided design format by injecting a paste with a robotic injector [25]. TGA of the composite PCL-MBG scaffolds revealed *ca* 40 wt-% of PCL mixed with MBGs.

Gelatin is a natural polymer that is very hydrophilic and highly degradable in aqueous medium. However, after being crosslinked with GA, gelatin decreases the latter degradability [26,27]. In the present study, gelatin was chosen as toughening polymer as it is a hydrolyzed form of collagen, which is a major constituent of the extracellular matrix of most tissues [28], a biocompatible and low antigenicity protein, and shows high bioabsorptivity in vivo [29,30]. Scaffolds were coated by immersion in 2.4 % gelatin crosslinked with GA (0.05 w/v%), giving rise to the following scaffolds: BL-GE, 4ZN-GE and 5ZN-GE, respectively.

*3.2 Characterization of raw MBGs and zinc-enriched MBGs*

Nitrogen adsorption–desorption isotherm and pore size distribution of BL, 4ZN and 5ZN MBGs, coated or not with GA crosslinked gelatin and loaded or unloaded with osteostatin, are shown in Figure 1. All curves can be identified as type IV isotherms, typical of mesoporous materials. They show type H1 hysteresis loops in the mesopore range, characteristic of cylindrical pores opened at both ends; the corresponding pore size distributions are also shown (Fig. 1, bottom). No significant variations in shape of the isotherms were found associated with gelatin coating or osteostatin loading in the scaffolds. Table 2 collects the textural properties, namely surface area, pore diameter and pore volume of the scaffolds. In general; the higher values of the textural properties were observed in the calcined samples. BL-GE, 4ZN-GE and 5ZN-GE scaffolds presented a slight decrease of surface area and pore volume due to gelatin coating, but these textural features remained high enough to host osteostatin molecules. Additional decreases of the specific surface area and pore volume were detected in BL-GE+OST, 4ZN-GE+OST and 5ZN-GE+OST scaffolds, confirming the efficacy of the peptide load into these scaffolds. It is also important to highlight that, when MBG was doped with increasing amounts of zinc, the mesoporous order decreased, losing its well-ordered mesoporosity. As a result, specific surface area and pore volume distribution decreased as the zinc content increased [25].

TGA was performed to determine the residual content of PCL from calcined scaffolds and the gelatin content in the coated ones (Table 2). The % weight of oxides in BL, 4ZN and 5ZN calcined scaffolds were determined by EDX analysis. The experimental values included in Table 2 fairly agree with the nominal ones that are also included between brackets.

Gelatin-coated MBG scaffolds were grinded and the resultant powders were characterized by SA-XRD. The presence of diffraction maxima in this region was indicative of mesoporous order (Figure 2A). However, differences between samples as a function of the ZnO content were observed. Thus, BL-GE displayed a sharp diffraction maximum at 2θ in the region of 1.0–1.4º, assigned to the (10) reflection, along with a poorly resolved peaks around 2.0 that can be assigned to the (11) reflection. These maxima were indexed on the basis of an ordered two-dimensional (2D) hexagonal structure (plane group *p6mm*) [6,31]. The intensity of the (10) maximum decreased associated with an increased ZnO in the scaffolds, indicating a progressive deterioration of the mesoporous structure in both 4ZN-GE and 5ZN-GE scaffolds.

To study the effectiveness and homogeneity of the biopolymer coating, FTIR spectroscopy and TGA studies were carried out. FITR spectra of GA crosslinked gelatin coated and uncoated scaffolds were collected (Figure 2B and inset). Spectra of native MBG scaffolds showed intense absorption bands at 1040 and 470 cm$^{-1}$ corresponding to the asymmetric bending vibration of the Si–O–Si bond, and a band at 800 cm$^{-1}$ attributed to symmetric stretching of the Si–O bond [32]. The GA crosslinked gelatin hydrogel presented –CH$_2$– groups and –NH$_2$– groups exhibiting bands between 3190 and 3440 cm,$^{-1}$ corresponding to C–H and N–H bonds. The stretching bands at 1600-1700 cm$^{-1}$ and 1500-1560 cm$^{-1}$ were assigned to C=O (Amide I) and N-C=O (Amide II) bonds, respectively. The deformation band at 1460-1600 cm$^{-1}$ was assigned to N–H, whereas 1510 and 1440 cm$^{-1}$ bands correspond to COO$^-$ bond [33, 34]. The C=N band at 1535 cm$^{-1}$ corresponds to crosslinked bond gelatin and the rest of bands to carboxyl and amino groups of the hydrogel. The amount of GA crosslinked gelatin film in BL-GE, 4ZN-GE and 5ZN-GE scaffolds was determined by TGA. The results revealed that the three types of MBG samples exhibited comparable gelatin contents, with weight losses of 6 ± 0.5% (Table 2). TEM studies of these MBG scaffolds confirmed the results obtained by XRD. Figure 3 shows TEM images corresponding to BL and 4ZN scaffolds obtained with the electron beam parallel to the mesoporous channels, indicating well-ordered hexagonally arranged mesostructures. TEM image of 5ZN scaffolds correspond to worm-like mesoporous structure [35].

SEM micrographs of the intact surface and taken after fracture of 4ZN and 4ZN-GE scaffolds, with and without GA crosslinked gelatin, are shown in Figure 4. The perpendicular view to the surface shows two types of channels of 700-1000 μm and 1000-1500 μm in diameter. Moreover, the designed architecture exhibited giant pores of around 400 μm between two adjacent rods in the cross-sectional fracture images after the biopolymer coating. In the 4ZN-GE scaffolds, it was possible to appreciate the GA crosslinked gelatin layer on the surface of the material. Higher magnification micrographs of the same region showed interconnected macropores in the range of 1–10 μm, as determined by mercury porosimetry. EDX spectra of the powders obtained by grinding the scaffolds showed the presence of Si, Ca, P and Zn in both 4ZN and 4ZN-GE samples (Fig. 4). Moreover, gelatin-containing scaffolds also showed the presence of C from the hydrogel. The $SiO_2$, $CaO$, $P_2O_5$ and $ZnO$ content of these MBGs obtained by EDX were in good agreement with the expected values.

Figure 5 exhibits the cumulative intrusion (A) and pore size distribution (B) obtained by mercury intrusion porosimetry of MBG powders, as well as of dried and rehydrated scaffolds. The dried BL-GE, 4ZN-GE and 5ZN-GE scaffolds were rehydrated after being soaked in water by 24 h, and then lyophilized in a freeze dryer LyoQuest Telstar (Tarrasa, Spain). This processing allowed us to forecast the effect of gelatin on the porosity of the scaffolds under in vitro and in vivo conditions. In fact, gelatin hydrogel swelling in aqueous media is expected to impact the fluid exchange inside the scaffold, with paramount consequence for biological processes because of the improvement of the mechanical properties and the diffusion processes. Mercury porosimetry measurements allowed us to determine how the scaffold porosity increased as a consequence of hydration. In this way, from the cumulative intrusion curves of Figure 5A, total porosity values were obtained: 38, 30 and 34 %, for freeze-dried BL-GE, 4ZN-GE and 5ZN-GE, respectively, which increased to 57, 51 and 42%, respectively, in the rehydrated scaffolds. This porosity increase took place mainly in the 15–600 μm pore range: 8% for BL-GE, 18% for 4ZN-GE and 10% for 5ZN-GE (Figure 5B). Thus, the gelatin coating should contribute to a better cell colonization.

$^{29}$Si (Figure 6A) and $^{31}$P (Figure 6B) solid state MAS NMR was carried out for investigating the network forming species on the different MBG scaffolds at atomic

level. $Q^2$, $Q^3$, and $Q^4$ represent the silicon atoms (denoted Si*) in (NBO)$_2$Si*–(OSi)$_2$, (NBO)Si*–(OSi)$_3$, and Si*(OSi)$_4$ (NBO = nonbonding oxygen), respectively. Table 3 shows the chemical shifts, deconvoluted peak areas, and silica network connectivity obtained from $^{29}$Si and $^{31}$P NMR spectroscopy of each MBG composition. The signals in the regions -110 ppm to -112 ppm and -100 ppm to -103 ppm come from $Q^4$ and $Q^3$, respectively; a resonance at approximately -92 ppm comes from $Q^2$ [36]. Using $^{29}$Si NMR spectroscopy, the MBG scaffolds were characterized by a high percentage of $Q^4$ and $Q^3$ species with a network connectivity of 3.62. These data are consistent with previous studies using similar MBG scaffolds [25]. The main ZnO structure in 4ZN-GE scaffolds is present as [ZnO$_4$] tetrahedra. These tetrahedra present a negative charge (2–), which justifies its attraction of Ca$^{2+}$ and Zn$^{2+}$ ions acting as network compensators of charge and not as network modifier cations. Accordingly, the number of NBO decreases with the increase in the % of $Q^4$ species and the decrease of $Q^3$ species (Figure 6A and Table 3).

It is important to note that MBG scaffolds, exhibiting higher $Q^3$ % and smaller $Q^2$ % than MBG disks, were submitted to two calcinations at 700 ºC, while MBG disks were only subjected to one calcination [24]. This network distribution should influence the in vitro behavior of the scaffolds compared to the disks. Moreover, the network connectivity of 4ZN-GE and 5ZN-GE scaffolds slightly increased until reach values similar to BL-GE scaffolds (Table 3). It contrasts with the network connectivity decrease, previously observed in 7% ZnO MBGs [25]. Tetrahedral symmetry of the $Q^4$ units in BL scaffolds respect to Zn-enriched scaffolds showed a more isotropic structure evidenced by the increase of the full width at half maximum (fwhm) parameter (See Table 3). In addition, the crystallinity of $Q^3$ and $Q^2$ was found to be slightly greater when ZnO was present in the samples.

$Q^0$ and $Q^1$ represent phosphorus atoms (denoted P*) in the PO$_4^{3-}$ species, (NBO)$_3$P*–(OP), (NBO)$_2$–P*–(OP)$_2$ (NBO denotes a nonbonding oxygen, relative to another P atom), respectively [24]. In Figure 6B, the single-pulse $^{31}$P MAS NMR spectra of the tested materials showed a maximum at ~2 ppm assigned to the $Q^0$ environment typical of amorphous orthophosphate. In addition to this signal from the main orthophosphate component, a minor $^{31}$P resonance was found to be present in all spectra. The latter signal was shown to shift from –8.0 ppm to –8.7 ppm with increasing ZnO% in MBGs. This resonance falls in the range of $Q^1$ tetrahedra and can be assigned to P–O–Si environment, as previously reported [34,35]. It is interesting to highlight that

ZnO introduction progressively causes a sightly decrease of $Q^1$ units percentage but a chemical shift increase in 4ZN and 5ZN scaffolds. This suggests a partial conversion of P–O–Si units into P–O–Zn units. P–O–Zn formation has previously been detected by Vallet-Regí *et al.* in other bioactive MBGs with analogous composition [24,25,38].

Changes in $Ca^{2+}$, P(V) and $Zn^{2+}$ ions released by the scaffolds, were measured soaking the scaffolds in MSCBM (Figure 7). For all the scaffolds, $Ca^{2+}$ and P(V) ion concentrations in medium underwent a sustained increase with time. For BL-GE scaffolds, the low concentration of $Ca^{2+}$ ions in the medium was attributed to its high in vitro bioactive behaviour. Thus, a part of the $Ca^{2+}$ ions released would precipitate as carbonate hydroxyapatite (CHA) on the scaffold surface being removed from the solution. In contrast, 4ZN-GE and 5ZN-GE scaffolds displayed slower bioactive response and, consequently, most part of the released $Ca^{2+}$ ions from these scaffolds remained in the liquid medium. Regarding phosphorous ions, 4ZN-GE and 5ZN-GE scaffolds showed lower concentration in the medium because of the precipitation of calcium and zinc phosphate [25,35]. In contrast, $Zn^{2+}$ release rate from these scaffolds was almost constant (about 1 ppm per day) during the first 5 d of soaking, but somewhat higher for 5ZN-GE compared to 4ZN-GE due to its higher ZnO content.

These results confirmed the medium accessibility to the inner part of the gelatin coated MBG scaffolds. Moreover, the fast hydrogel swelling (30 min) enhanced $Ca^{2+}$, P(V) and $Zn^{2+}$ ions release, compared with the uncoated scaffolds [25]. This is due to hydration, occurring after GA cross-linked gelatin scaffold swelling. In an aqueous environment, hydrogel allows water exchange and, subsequently, ion release increases. No different behaviour regarding the ion release profile was observed in the osteostatin-loaded scaffolds (data not shown). Of note, $Zn^{2+}$ released from 4ZN-GE and 5ZN-GE scaffolds was larger than that of 4ZN and 5ZN as dense disks of the same composition as the scaffolds [24]. This could be explained because of the high network connectivity of these dense glasses; $Zn^{2+}$ ions seem to mainly behave as network formers, more difficult to release to the surrounding medium. Moreover, the open interconnected macroporosity and the gelatin coating of the scaffolds facilitate ions release, compared to the disks. This makes the scaffolds an improved material for biological applications.

*3.3 Rehydration effect of the porous scaffolds*

The polar groups of gelatin make it a hydrophilic and absorbent polymer [37,38]. To investigate the behavior of gelatin-coated MBG scaffolds in body fluids, rehydration assays were carried out. Figure 8A illustrates the swelling ratio of the different scaffolds along time; the %W curves show a fast increase of water absorption in the first 15 min, followed by a slight increment during the following 24 h. BL-GE, 4ZN-GE and 5ZN-GE scaffolds absorb between 120-140 wt-% of water by respect to their initial dry mass. This high grade of hydration could be explained by the presence of 6-7 wt. % GA crosslinked gelatin in the scaffolds (Table 2) combined with their high interconnected macroporosity (Figures 4 and 5). The swelling capability and pore size enlargement displayed by all the scaffolds was similar due to their alike macropore size distribution and pores interconnectivity [39,40].

*3.4 Osteostatin loading and release in MBG scaffolds*

We found that the mean retention of osteostatin by the tested scaffolds after 24 h of loading was: 47 % (BL-GE+OST), 52 % (4ZN-GE+OST) and 52% (5ZN-GE+OST), equivalent to 0.52, 0.71 or 0.62 µg osteostatin/g scaffold, respectively. These loaded materials released (mean): 56 % (BL-GE+OST), 52% (4ZN-GE+OST) or 59% (5ZN-GE+OST) of loaded peptide to the surrounding medium within 1 h; 90% (BL-GE+OST), 87% (4ZN-GE+OST) and 90% (5ZN-GE+OST) at 24 h; and virtually 100 % in all these scaffolds at 96 h (Figure 8B). This indicates that minimal osteostatin retention occurs in these biomaterials. Osteostatin uptake by MBG scaffolds was of the same magnitude order in µg/g than previously observed in MBG disks [22,24]. However, for the reasons that will be given later was somewhat higher in the scaffolds. Interestingly in this regard, our group previously demonstrated that this peptide, at concentrations in the sub-nM range, was efficient to increase osteoblast function in vitro [23,41,42]. The release mechanism of osteostatin is assumed to be diffusion through the mesopores [43]. Therefore, that silica matrix is insoluble at pH 7.4, the release of this peptide to the surrounding solution could be described by a deviation from the theoretical first-order behaviour in the Noyes–Whitney equation with the introduction of an empirical non-ideally factor $\delta$ [44,45]:

$$w_t/w_0 = A(1-\exp k_1 \cdot t)^{\delta}$$

$W_t$ stands for the peptide mass released at time t; $w_0$ represents the maximum initial mass of the peptide inside the pores; A is the maximum amount of peptide released; and $k_1$ is the release rate constant, which is independent of peptide concentration and gives information about the solvent accessibility and the diffusion coefficient through mesoporous channels. The peptide release profiles showed a deviation from the theoretical first-order behaviour described by the equation above (Figure 8B): faster release within 24 h, reaching a stationary phase after 48 h. This deviation could be due to several factors, such as the volume of the peptide, the distortion of the mesopore channels and/or the release of peptide molecules adsorbed on the external surface of the scaffold matrices. Values for δ range between 1, for materials that follows first-order kinetics, and 0 for those rapidly releasing the load located at pore entrances. Table 4 shows the release parameters by plotting our data using this semi-empirical first-order model. According to this model, δ gives an idea of the accuracy of this approximation. This parameter was found to be similar for all the scaffolds studied, indicating that osteostatin molecules are released from the internal and external surface of the mesopore structure of each loaded glass (Table 4). It is worth noticing that peptide load (µg/g) is somewhat higher in MBG scaffolds than in analogous MBG disks [24], due to more accessible mesopore structure by their hierarchical channels and macroporosity.

*3.5 In vitro cytocompatibility assays*

*3.5.1 Cell morphological studies by confocal laser scanning microscopy and cell growth*

3D confocal microscopy studies were carried out on the cylindrical scaffolds studied after hMSCs seeding (Figure 9A). The geometrical center of each scaffold (6 x10 mm$^2$) was scanned to determine the cell morphology after 5 d, using Atto 565–phalloidin as fluorescence probe of F-actin microfilaments. In this case, DAPI fluorostaining was not used because it also dyes the gelatin coating of the scaffolds. Actin is the most important interconnected filament protein in the cytoskeleton, performing critical cell functions, including cell shape maintenance, cell movement as well as intracellular trafficking [46]. We found that hMSCs presented their typical spindle-shaped morphology and were adequately spread on the scaffold surface, indicating the biocompatibility of every type of scaffolds tested. Furthermore, cell clusters interconnected at different depths in the scaffold were also observed as evidence of optimal internalization and intercellular communication in the 4ZN-GE scaffolds**.**

We next studied and compared the osteogenic activity conferred by zinc and osteostatin to these scaffolds as carriers of hMSCs. Human MSC growth was increased in zinc-containing scaffolds, compared to BL scaffolds, at 5 and 7 d; and this positive effect was significantly higher in 4ZN-GE and 5ZN-GE materials loaded with osteostatin (BL-GE+OST, 4ZN-GE+OST and 5ZN-GE+OST) at both time periods (Figure 9B), indicating the synergistic effect between both factors in this context. Consistent with this finding, hMSC growth and spreading occurred on both the surface and inside of the zinc-enriched and native MBG scaffolds with osteostatin load (Figure 9A).

*3.5.2 Cytotoxicity assay: Lactate dehydrogenase (LDH) activity*

None of the tested materials induced significant cell death assessed by cell LDH release; but 5ZN-GE scaffolds exhibited a (not significant) tendency to increase LDH levels, which was counteracted by osteostatin in this type of scaffold at day 10 of cell culture (Figure 10). This would be due to a higher degradation of 5ZN-GE scaffold compared to the other scaffolds studied, and it might account for the lower cell growth observed in this scaffold at 5 and 7 d. Lozano et al. [23] previously demonstrated in vitro that exposure of the pre-osteoblast MC3T3-E1 cell line to osteostatin-loaded ordered mesoporous SBA-15 disks stimulated cell growth and osteoblastic differentiation. Subsequent studies showed that SBA-15 based ceramics containing osteostatin as implants are biocompatible and induce recruitment and activation of osteoprogenitors to promote bone regeneration in osteoporotic rabbits [31]. Previous studies also have shown that 5% appears to be the maximum amount of ZnO to be incorporated into certain types of bioactive glasses causing no cytotoxicity while enhancing osteoblast cell growth [47–49]. In the same line, Vallet-Regí's group recently demonstrated that MBGs doped with 7% ZnO were cytotoxic for MC3T3-E1 cells [25].

*3.5.3 Cell differentiation: RUNX2 and ALP gene expression*

We next evaluated the capacity of these scaffolds, loaded or not with osteostatin, to affect hMSC differentiation by evaluating the gene expression of two osteoblast differentiation markers, RUNX2 and ALP. RUNX2 is an early osteogenic differentiation marker that directs cell progenitors toward the osteoblastic lineage [50]. We found that both 4ZN-GE and 5ZN-GE scaffolds were inefficient to affect RUNX2 gene expression in hMSCs cultures at day 7, but the simultaneous presence of osteostatin in these scaffolds increased this expression at this time (Figure 11A). ALP is considered as a potential key biochemical marker of osteogenic differentiation, which acts to cleave phosphate groups from different substrates during osteogenesis [51]. Here, we found that 4ZN-GE+Ost and 5ZN-GE+Ost scaffolds significantly increased the gene expression of ALP at day 7 of hMSC culture, compared to osteostatin-unloaded scaffolds (Figure 11B).

It has previously been suggested that the putative benefits of adding $Zn^{2+}$ to a bioactive glass would be impaired by $Zn^{2+}$-promoted material degradation. In fact, no clear data have been presented to date to confirm the osteogenic effect of zinc decorating bioactive silicate glasses [52,53]. In our recent studies [24], osteostatin load was found to confer zinc-enriched MBG disks the ability to stimulate ALP activity and matrix mineralization in cultured MC3T3-E1 cells in an osteoblast differentiation medium. In the present research using hMSCs, the combination of zinc and osteostatin into 4ZN-GE+OST and 5ZN-GE+OST scaffolds significantly increased cell growth and thorough cell colonization of these materials compared to BL-GE scaffolds. Furthermore, zinc and osteostatin, in conjunction but not independently, were shown to induce the gene expression of the osteoblastic differentiation markers RUNX2 and ALP in hMSCs in the absence of a promoting differentiation medium (Figure 11).

*3.6 Synergistic effect of zinc and osteostatin on hMSC growth and osteogenic differentiation*

The underlying mechanism of the observed synergistic effect of zinc and osteostatin on hMSC growth and osteogenic differentiation is unknown at present. Zinc has important roles in cell signal transduction and gene expression, and as a catalytic component of a variety of mammalian enzymes; however, its putative effects on MSCs have been poorly investigated [54]. A previous study, using sol-gel bioactive glass granules containing 2-5% zinc has shown a stimulatory effect of this cation on osteogenic differentiation -based on ALP activity- of adult rat MSCs [55].

On the other hand, another study reported that zinc ions, at sub or supra nM concentrations, inhibited osteogenic differentiation of mouse MSCs [56]. Moreover, another report has shown that ZnT7, a member of the zinc transporter family SLC30A, can act as an inhibitor during dexamethasone-induced differentiation of rat MSCs; an effect which appears to be dependent in part on targeting extracellular signal-regulated kinase (ERK) signaling pathway [57]. In addition, a variety of in vitro and in vivo studies demonstrate the osteogenic effects of osteostatin [20,21,58]. Of note, one of these studies has reported that osteostatin synergistically interacts with Si-HA-coupled fibroblast growth factor-2 to induce ALP and matrix mineralization in MC3T3-E1 and primary human osteoblastic cells by an ERK-dependent mechanism [59]. A possible interaction between Zn ions and osteostatin through ERK activation, or any alternative intracellular mechanism(s), to improve the osteogenic efficacy of our scaffolds is an attractive hypothesis that deserves further studies.

The osteoblastic lineage is known to be regulated by various molecules including systemic and local peptide factors. As in other tissues, in which peptides can act as cell mediators, cell surface peptidases can modulate their activity. In this regard, zinc-metallopeptidase neprilysin, a cell surface peptidase present in all the cells of the osteoblastic lineage (preosteoblasts, mature osteoblasts and osteocytes) [60,61], is upregulated in hMSCs during osteogenic differentiation [62]. Also of interest, this zinc-metallopeptidase can cleave several osteogenic peptides, including native PTHrP 107-139 (containing the osteostatin sequence 107-111). In this scenario, it could be speculated that endogenous PTHrP, which increases with osteoblast differentiation [17,18], might collaborate with $Zn^{2+}$ to promote such differentiation. Although further studies are waiting to confirm this hypothesis, the synergistic effect observed with our zinc-containing MBG scaffolds after priming with osteostatin in the present context supports this rationale.

In any event, in the present study, we provide the first evidence that the combination of zinc and osteostatin in MBG scaffolds can promote cell proliferation and differentiation of hMSCs without classical differentiation inductors, suggesting the potential of this approach in bone tissue engineering applications.

**4. Conclusions**

This study reports the synthesis and design, using a rapid prototyping technique, of meso-macroporous MBG 3D-scaffolds coated with GA-crosslinked gelatin, and enriched with zinc and osteostatin. These 3-D hierarchical interconnected macroporous scaffolds show excellent cellular internalization and an outstanding hMSCs response in terms of cell adhesion, growth and osteogenic differentiation. The composition and 3D-architecture of GA crosslinked gelatin coating the hierarchical porous of MBG scaffolds facilitate the controlled release of both zinc and osteostatin, which positively affects hMSCs development and their osteogenic differentiation in medium without classical differentiation inductors. This system allowed us to disclose, for the first time, a synergistic effect of zinc and osteostatin to enhance hMSC cell growth and osteogenic differentiation, suggesting its potential use in bone tissue engineering applications.

**Acknowledgements**

This study was supported by research grants from Instituto de Salud Carlos III (PI15/00978) project co-financed with the European Union FEDER funds, the European Research Council (ERC-2015-AdG) Advanced Grant Verdi-Proposal No.694160 and MINECO MAT2015-64831-R project.

**References**

[1] S. Bose, M. Roy, A. Bandyopadhyay, Recent advances in bone tissue engineering scaffolds, Trends Biotechnol. 30 (2012) 546–554.

[2] S.L. Wu, X.M. Liu, K.W.K. Yeung, C.S. Liu, X.J. Yang, Biomimetic porous scaffolds for bone tissue engineering, Mater. Sci. Eng. R-Rep. 80 (2014) 1–36.

[3] A.J. Salinas, P. Esbrit, M. Vallet-Regi, A tissue engineering approach based on the use of bioceramics for bone repair, Biomater. Sci. 1 (2013) 40–51.

[4] X.X. Yan, C.Z. Yu, X.F. Zhou, J.W. Tang, D.Y. Zhao, Highly ordered mesoporous bioactive glasses with superior in vitro bone-forming bioactivities, Angew. Chem.-Int. Edit. 43 (2004) 5980–5984.

[5] M. Vallet-Regi, A.J. Salinas, D. Arcos, Tailoring the Structure of Bioactive Glasses: From the Nanoscale to Macroporous Scaffolds, Int. J. Appl. Glass Sci. 7 (2016) 195–205.

[6] A. Lopez-Noriega, D. Arcos, I. Izquierdo-Barba, Y. Sakamoto, O. Terasaki, M. Vallet-Regi, Ordered mesoporous bioactive glasses for bone tissue regeneration, Chem. Mat. 18 (2006) 3137–3144.

[7] G. Jell, M.M. Stevens, Gene activation by bioactive glasses, Journal of Materials Science-Materials in Medicine 17 (2006) 997–1002.

[8] L.L. Hench, Genetic design of bioactive glass, J. Eur. Ceram. Soc. 29 (2009) 1257–1265.

[9] S. Kaya, M. Cresswell, A.R. Boccaccini, Mesoporous silica-based bioactive glasses for antibiotic-free antibacterial applications, Mater. Sci. Eng. C-Mater. Biol. Appl. 83 (2018) 99–107.

[10] J. Chou, J. Hao, H. Hatoyama, B. Ben-Nissan, B. Milthorpe, M. Otsuka, Effect of biomimetic zinc-containing tricalcium phosphate (Zn-TCP) on the growth and osteogenic differentiation of mesenchymal stem cells, J. Tissue Eng. Regen. Med. 9 (2015) 852–858.

[11] Y.Q. Yu, G.D. Jin, Y. Xue, D.H. Wang, X.Y. Liu, J. Sun, Multifunctions of dual Zn/Mg ion co-implanted titanium on osteogenesis, angiogenesis and bacteria inhibition for dental implants, Acta Biomater. 49 (2017) 590–603.

[12] S.D. Gittard, J.R. Perfect, N.A. Monteiro-Riviere, W. Wei, C.M. Jin, R.J. Narayan, Assessing the antimicrobial activity of zinc oxide thin films using disk diffusion and biofilm reactor, Appl. Surf. Sci. 255 (2009) 5806–5811.

[13] Z.Z. Chu, T.R. Zhao, L. Li, J. Fan, Y.Y. Qin, Characterization of Antimicrobial Poly (Lactic Acid)/Nano-Composite Films with Silver and Zinc Oxide Nanoparticles, Materials 10 (2017) 13.

[14] J.Y. Tan, C.K. Chua, K.F. Leong, Fabrication of channeled scaffolds with ordered array of micro-pores through microsphere leaching and indirect Rapid Prototyping technique, Biomed. Microdevices 15 (2013) 83–96.

[15] M. Cicuendez, I. Izquierdo-Barba, S. Sanchez-Salcedo, M. Vila, M. Vallet-Regi, Biological performance of hydroxyapatite-biopolymer foams: In vitro cell response, Acta Biomater. 8 (2012) 802–810.

[16] D.M.R. Gibbs, C.R.M. Black, J.I. Dawson, R.O.C. Oreffo, A review of hydrogel use in fracture healing and bone regeneration, J. Tissue Eng. Regen. Med. 10 (2016) 187–198.

[17] P. Esbrit, M.J. Alcaraz, Current perspectives on parathyroid hormone (PTH) and PTH-related protein (PTHrP) as bone anabolic therapies, Biochem. Pharmacol. 85 (2013) 1417–1423.

[18] N.S. Datta, A.B. Abou-Samra, PTH and PTHrP signaling in osteoblasts, Cell. Signal. 21 (2009) 1245–1254.

[19] A.J. Fenton, B.E. Kemp, R.G. Hammonds, K. Mitchelhill, J.M. Moseley, T.J. Martin, G.C. Nicholson, a potent inhibitor of osteoclastic bone-resorption within a highly conserved pentapeptide region of parathyroid hormone-related protein - PTHrP 107-111, Endocrinology 129 (1991) 3424–3426.

[20] D. Lozano, L. Fernandez-de-Castro, S. Portal-Nunez, A. Lopez-Herradon, S. Dapia, E. Gomez-Barrena, P. Esbrit, The C-terminal fragment of parathyroid hormone-related peptide promotes bone formation in diabetic mice with low-turnover osteopaenia, Br. J. Pharmacol. 162 (2011) 1424–1438.

[21] J. Cornish, K.E. Callon, C. Lin, C.R. Xiao, J.M. Moseley, I.R. Reid, Stimulation of osteoblast proliferation by C-terminal fragments of parathyroid hormone-related protein, J. Bone Miner. Res. 14 (1999) 915–922.

[22] J.A. Ardura, S. Portal-Nunez, D. Lozano, I. Gutierrez-Rojas, S. Sanchez-Salcedo, A. Lopez-Herradon, F. Mulero, M.L. Villanueva-Penacarrillo, M. Vallet-Regi, P. Esbrit, Local delivery of parathyroid hormone-related protein-derived peptides coated onto a hydroxyapatite-based implant enhances bone regeneration in old and diabetic rats, J. Biomed. Mater. Res. Part A 104 (2016) 2060–2070.

[23] D. Lozano, M. Manzano, J.C. Doadrio, A.J. Salinas, M. Vallet-Regi, E. Gomez-Barrena, P. Esbrit, Osteostatin-loaded bioceramics stimulate osteoblastic growth and differentiation, Acta Biomater. 6 (2010) 797–803.

[24] R. Perez, S. Sanchez-Salcedo, D. Lozano, C. Heras, P. Esbrit, M. Vallet-Regi, A.J. Salinas, Osteogenic Effect of ZnO-Mesoporous Glasses Loaded with Osteostatin, Nanomaterials 8 (2018) 592.

[25] S. Sanchez-Salcedo, S. Shruti, A.J. Salinas, G. Malavasi, L. Menabue, M. Vallet-Regi, In vitro antibacterial capacity and cytocompatibility of $SiO_2$-CaO-$P_2O_5$ meso-macroporous glass scaffolds enriched with ZnO, J. Mat. Chem. B 2 (2014) 4836−4847.

[26] H.C. Yang SH, Wang KC, Hou SM, Lin FH., Tricalcium phosphate and glutaraldehyde crosslinked gelatin incorporating bone morphogenetic protein-a viable scaffold for bone tissue engineering, J Biomed Mater Res B Appl Biomater. 74 (2005 ) 468−475.

[27] P. Verma, V. Verma, P. Ray, A.R. Ray, Agar-gelatin hybrid sponge-induced three-dimensional in vitro 'liver-like' HepG2 spheroids for the evaluation of drug cytotoxicity, J. Tissue Eng. Regen. Med. 3 (2009) 368−376.

[28] O. Mahony, O. Tsigkou, C. Ionescu, C. Minelli, L. Ling, R. Hanly, M.E. Smith, M.M. Stevens, J.R. Jones, Silica-Gelatin Hybrids with Tailorable Degradation and Mechanical Properties for Tissue Regeneration, Adv. Funct. Mater. 20 (2010) 3835−3845.

[29] F. Perut, E.B. Montufar, G. Ciapetti, M. Santin, J. Salvage, T. Traykova, J.A. Planell, M.P. Ginebra, N. Baldini, Novel soybean/gelatine-based bioactive and injectable hydroxyapatite foam: Material properties and cell response, Acta Biomater. 7 (2011) 1780−1787.

[30] J. Heino, The collagen family members as cell adhesion proteins, Bioessays 29 (2007) 1001−1010.

[31] D. Lozano, C.G. Trejo, E. Gomez-Barrena, M. Manzano, J.C. Doadrio, A.J. Salinas, M. Vallet-Regi, N. Garcia-Honduvilla, P. Esbrit, J. Bujan, Osteostatin-loaded onto mesoporous ceramics improves the early phase of bone regeneration in a rabbit osteopenia model, Acta Biomater. 8 (2012) 2317−2323.

[32] C.J. Brinker, Sol-gel science: the physics and chemistry of sol-gel processing, , Academic Press, Boston, 1990, pp. 912.

[33] R. Shi, H. Geng, M. Gong, J. Ye, C. Wu, X. Hu, L. Zhang, Long-acting and broad-spectrum antimicrobial electrospun poly (ε-caprolactone)/gelatin micro/nanofibers for wound dressing, J Colloid Interface Sci. 509 (2018), 275-284.

[34] M. Xu, L. Wei, Y. Xiao, H. Bi, H. Yang, Y. Du, Physicochemical and functional properties of gelatin extracted from Yak skin, Int J Biol Macromol. 95 (2017), 1246-1253

[35] S. Shruti, A.J. Salinas, G. Lusvardi, G. Malavasi, L. Menabue, M. Vallet-Regi, Mesoporous bioactive scaffolds prepared with cerium-, gallium- and zinc-containing glasses, Acta Biomater. 9 (2013) 4836−4844.

[36] E. Leonova, I. Izquierdo-Barba, D. Arcos, A. Lopez-Noriega, N. Hedin, M. Vallet-Regi, M. Eden, Multinuclear solid-state NMR studies of ordered mesoporous bioactive glasses, J. Phys. Chem. C 112 (2008) 5552−5562.

[37] A. Garcia, M. Cicuendez, I. Izquierdo-Barba, D. Arcos, M. Vallet-Regi, Essential Role of Calcium Phosphate Heterogeneities in 2D-Hexagonal and 3D-Cubic SiO2-CaO-P2O5 Mesoporous Bioactive Glasses, Chem. Mat. 21 (2009) 5474−5484.

[38] L. Linati, G. Lusvardi, G. Malavasi, L. Menabue, M.C. Menziani, P. Mustarelli, U. Segre, Qualitative and quantitative structure-property relationships analysis of multicomponent potential bioglasses, J. Phys. Chem. B 109 (2005) 4989−4998.

[39] B. Bageshlooyafshar, S. Vakilian, F. Rafeie, R. Ramezanifard, R. Rahchamani, A. Mohammadi-Sangcheshmeh, Y. Mostafaloo, E. Seyedjafari, Zinc silicate mineral-coated

scaffold improved in vitro osteogenic differentiation of equine adipose-derived mesenchymal stem cells., Res Vet Sci. (2017) pii: S0034-5288(16)30814-1. doi: 10.1016/j.rvsc.2017.09.015.

[40] J.W. W. Friess, Handbook of Porous Solids, Biomedical applications, Wiley-VCH, Germany, 2002. pp. 554–562.

[41] D. Lozano, S. Sanchez-Salcedo, S. Portal-Nunez, M. Vila, A. Lopez-Herradon, J.A. Ardura, F. Mulero, E. Gomez-Barrena, M. Vallet-Regi, P. Esbrit, Parathyroid hormone-related protein (107-111) improves the bone regeneration potential of gelatin-glutaraldehyde biopolymer-coated hydroxyapatite, Acta Biomater. 10 (2014) 3307−3316.

[42] D. Lozano, M.J. Feito, S. Portal-Nunez, R.M. Lozano, M.C. Matesanz, M.C. Serrano, M. Vallet-Regi, M.T. Portoles, P. Esbrit, Osteostatin improves the osteogenic activity of fibroblast growth factor-2 immobilized in Si-doped hydroxyapatite in osteoblastic cells, Acta Biomater. 8 (2012) 2770−2777.

[43] F. Balas, M. Manzano, P. Horcajada, M. Vallet-Regi, Confinement and controlled release of bisphosphonates on ordered mesoporous silica-based materials, J. Am. Chem. Soc. 128 (2006) 8116−8117.

[44] J. Crank, The Mathematics of Diffusion, 2nd ed., Oxford University Press, Oxford, 1975.

[45] M. Manzano, V. Aina, C.O. Arean, F. Balas, V. Cauda, M. Colilla, M.R. Delgado, M. Vallet-Regi, Studies on MCM-41 mesoporous silica for drug delivery: Effect of particle morphology and amine functionalization, Chem. Eng. J. 137 (2008) 30–37.

[46] D.E. Ingber, The riddle of morphogenesis - A question of solution chemistry or molecular cell engineering, Cell 75 (1993) 1249–1252.

[47] S. Haimi, G. Gorianc, L. Moimas, B. Lindroos, H. Huhtala, S. Raty, H. Kuokkanen, G.K. Sandor, C. Schmid, S. Miettinen, R. Suuronen, Characterization of zinc-releasing three-dimensional bioactive glass scaffolds and their effect on human adipose stem cell proliferation and osteogenic differentiation, Acta Biomater. 5 (2009) 3122−3131.

[48] V. Salih, A. Patel, J.C. Knowles, Zinc-containing phosphate-based glasses for tissue engineering, Biomed. Mater. 2 (2007) 11–20.

[49] V. Aina, G. Malavasi, A.F. Pla, L. Munaron, C. Morterra, Zinc-containing bioactive glasses: Surface reactivity and behaviour towards endothelial cells, Acta Biomater. 5 (2009) 1211−1222.

[50] H. Drissi, Q.Y. Luc, R. Shakoori, S.C.D. Lopes, J.Y. Choi, A. Terry, M. Hu, S. Jones, J.C. Neil, J.B. Lian, J.L. Stein, A.J. Van Wijnen, G.S. Stein, Transcriptional autoregulation of the bone related CBFA1/RUNX2 gene, J. Cell. Physiol. 184 (2000) 341−350.

[51] J.E. Coleman, Structure and mechanism of alkaline-phosphatase, Annu. Rev. Biophys. Biomolec. Struct. 21 (1992) 441−483.

[52] A. Hoppe, N.S. Guldal, A.R. Boccaccini, A review of the biological response to ionic dissolution products from bioactive glasses and glass-ceramics, Biomaterials 32 (2011) 2757−2774.

[53] G. Lusvardi, G. Malavasi, L. Menabue, M.C. Menziani, A. Pedone, U. Segre, V. Aina, A. Perardi, C. Morterra, F. Boccafoschi, S. Gatti, M. Bosetti, M. Cannas, Properties of zinc releasing surfaces for clinical applications, J. Biomater. Appl. 22 (2008) 505−526.

[54] C.W. Levenson, D. Morris, Zinc and neurogenesis: making new neurons from development to adulthood, Advances Nut. 2, (2011) 96–100.

[55] S. A.Oh, S.H.Kim, J.E.Won, J.J.Kim, U.S. Shin, H.W. Kim, Effects on growth and osteogenic differentiation of mesenchymal stem cells by the zinc-added sol-gel bioactive glass granules, J. Tissue Eng. (2011) Jan 12; 2010: 475260.

[56] T. Wang, J.C. Zhang, Y. Chen, P.G. Xiao, M.S. Yang, Effect of zinc ion on the osteogenic and adipogenic differentiation of mouse primary bone marrow stromal cells and the adipocytic trans-differentiation of mouse primary osteoblasts, J. Trace Elem. Med. Biol., 21 (2007) 84–91.

[57] Y. Liu, F. Yan, W.L. Yang, X.F. Lu, W.B. Wang, Effects of zinc transporter on differentiation of bone marrow mesenchymal stem cells to osteoblasts, Biol. Trace Element Res., 154 (2013) 234–243.

[58] A.J Salinas, P. Esbrit, M. Vallet-Regí, A tissue engineering approach based on the use of bioceramics for bone repair. Biomater. Sci. 1 (2013) 40–51.

[59] D. Lozano, M.J. Feito, S. Portal-Núñez, R.M. Lozano, M.C. Matesanz, M.C. Serrano, M. Vallet-Regi, M.T. Portolés, Esbrit P, Osteostatin improves the osteogenic activity of fibroblast growth factor-2 immobilized in Si-doped hydroxyapatite in osteoblastic cells. Acta Biomater. 8 (2012) 2770–2777.

[60] D.N. Crine P, Boileau G Cell-Surface Peptidases in Health and Disease, 1st ed., Oxford, 1997, pp. 1266–1274.

[61] A.F. Ruchon, M. Marcinkiewicz, K. Ellefsen, A. Basak, J. Aubin, P. Crine, G. Boileau, Cellular localization of neprilysin in mouse bone tissue and putative role in hydrolysis of osteogenic peptides, J. Bone Miner. Res. 15 (2000) 1266–1274.

[62] C. Graneli, A. Thorfve, U. Ruetschi, H. Brisby, P. Thomsen, A. Lindahl, C. Karlsson, Novel markers of osteogenic and adipogenic differentiation of human bone marrow stromal cells identified using a quantitative proteomics approach, Stem Cell Res. 12 (2014) 153–165.

**FIGURE CAPTIONS**

**Figure 1.** $N_2$ adsorption-desorption isotherms and the corresponding pore size distributions of the scaffolds before and after being coated with gelatin (GE) and before and after being loaded with osteostatin (OST).

**Figure 2.** DRX patterns (A) and FITR spectra (B) of BL and zinc-doped scaffolds (4ZN and 5ZN) before and after being coated with GA crosslinked gelatin. *A magnification of (B).

**Figure 3.** TEM images of BL, 4ZN and 5ZN showing the first two a 2D hexagonal mesoporous structure and the last one a worm-like mesoporous structure.

**Figure 4.** SEM images of fracture and surface of 4ZN and 4ZN-GE scaffolds. At the bottom, the corresponding EDX spectra.

**Figure 5.** (A) Cumulative Hg intrusion volume as a function of the pore diameter. (B) Pore size distribution of BL-GE, 4ZN-GE and 5ZN-GE powders (P-), dried and rehydrated scaffolds (H-).

**Figure 6.** Solid-state (A) $^{29}Si$ and (B) $^{31}P$ single-pulse MAS-NMR spectra of BL-GE, 4ZN-GE and 5ZN-GE scaffolds. $Q^n$ unit areas were calculated by Gaussian line-shape deconvolution and displayed in green.

**Figure 7.** Evolution with time of the ionic concentration of calcium, phosphorus and zinc after soaking BL-GE, 4ZN-GE and 5ZN-GE scaffolds between 6 h and 10 d in complete medium.

**Figure 8.** (A) Swelling ratio (%*W*) of BL-GE, 4ZN-GE and 5ZN-GE scaffolds as a function of incubation time. (B) Osteostatin release from MBGs scaffolds between 1 h and 4 d in PBS. Points to trace the curves are the means of three independent measurements per time period. SEM values (representing a coefficient of variation of <5% for each point) are omitted for simplification.

**Figure 9.** (A) hMSCs cultured onto the BL-GE+OST, 4ZN-GE+OST and 5ZN-GE+OST scaffolds, by 5 d, were observed with confocal laser scanning microscopy. Atto 565–phalloidin was used as fluorescence probes to determine the cell morphology, Atto 565–phalloidin staining the F-actin microfilaments to visualize the cytoskeleton

indicating cell viability. Blue autofluorescence of the gelatin/MBG is also evident. The reflection of the scaffold material was visualized in green. (For interpretation of the references to colour in this figure legend, the reader is referred to the web version of this article). (B) Proliferation of MC3T3-E1 preosteblast-like cells as a function of culture time onto BL, 4ZN and 5ZN scaffolds with and without osteostatin at 5 d and 7 d.

**Figure 10**. LDH of hMSCs in the presence of BL-GE, 4ZN-GE and 5ZN-GE scaffolds loaded or not with osteostatin after 10 d of cell culture. Results are means ± SEM of three measurements in triplicate.

**Figure 11**. Runx2 (A) and ALP (B) mRNA levels (measured by real-time PCR) in hMSCs in the presence of BL-GE, 4ZN-GE and 5ZN-GE scaffolds loaded or not osteostatin at 7 d of culture. Results are mean ± SE (n = 5). *$p < 0.05$ vs the corresponding unloaded scaffold.

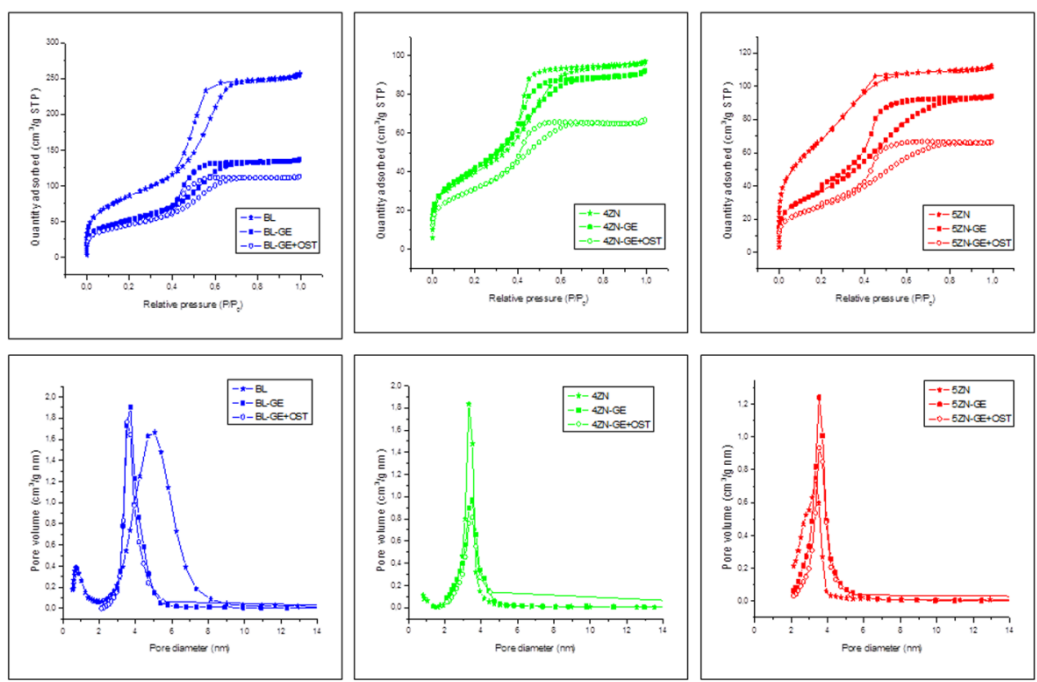

Figure 1

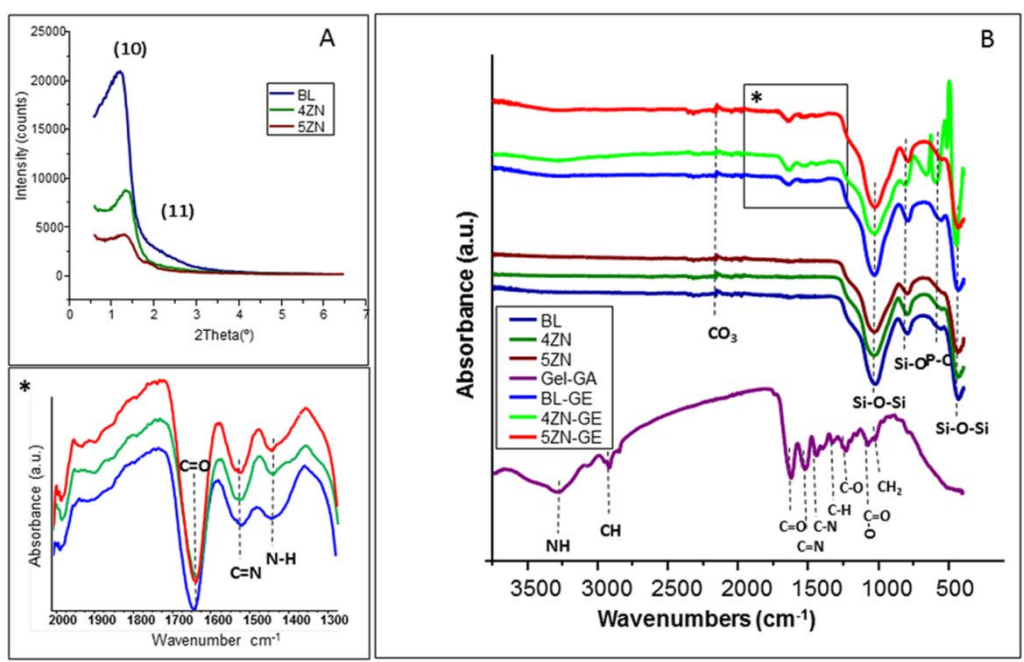

Figure 2

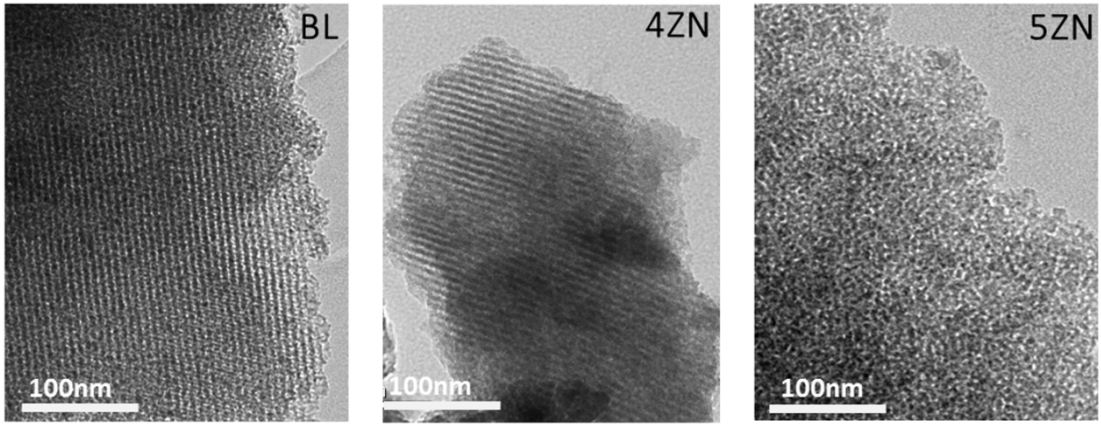

Figure 3

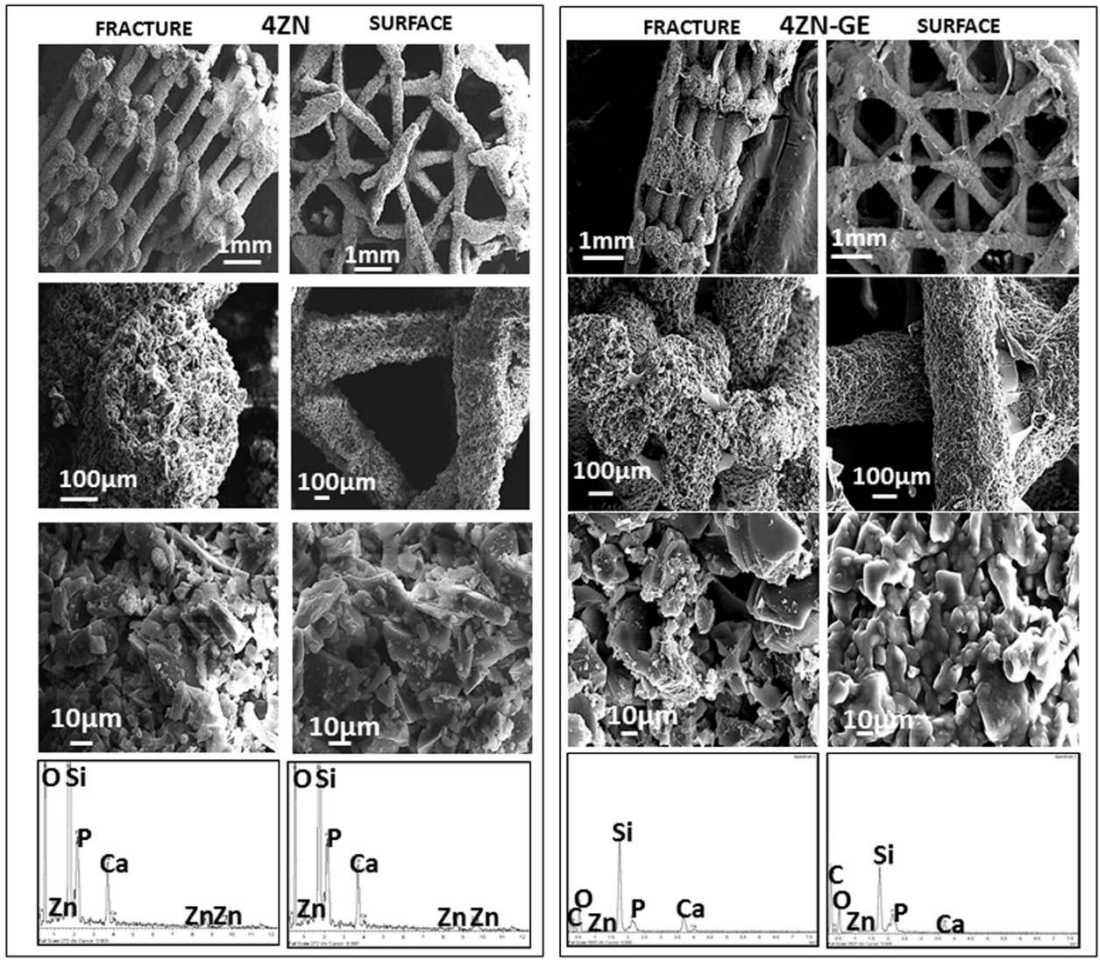

Figure 4

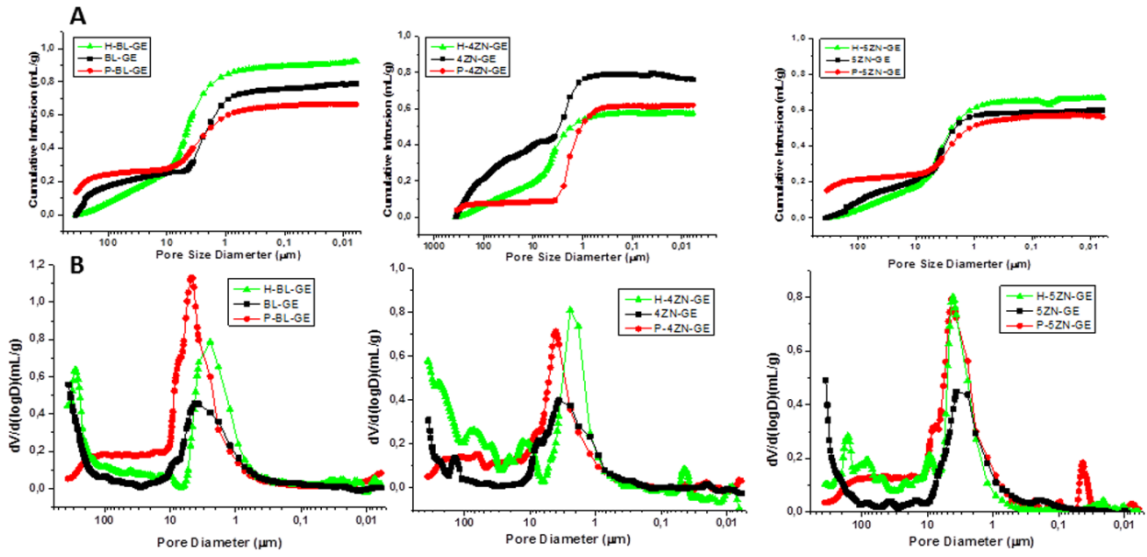

Figure 5

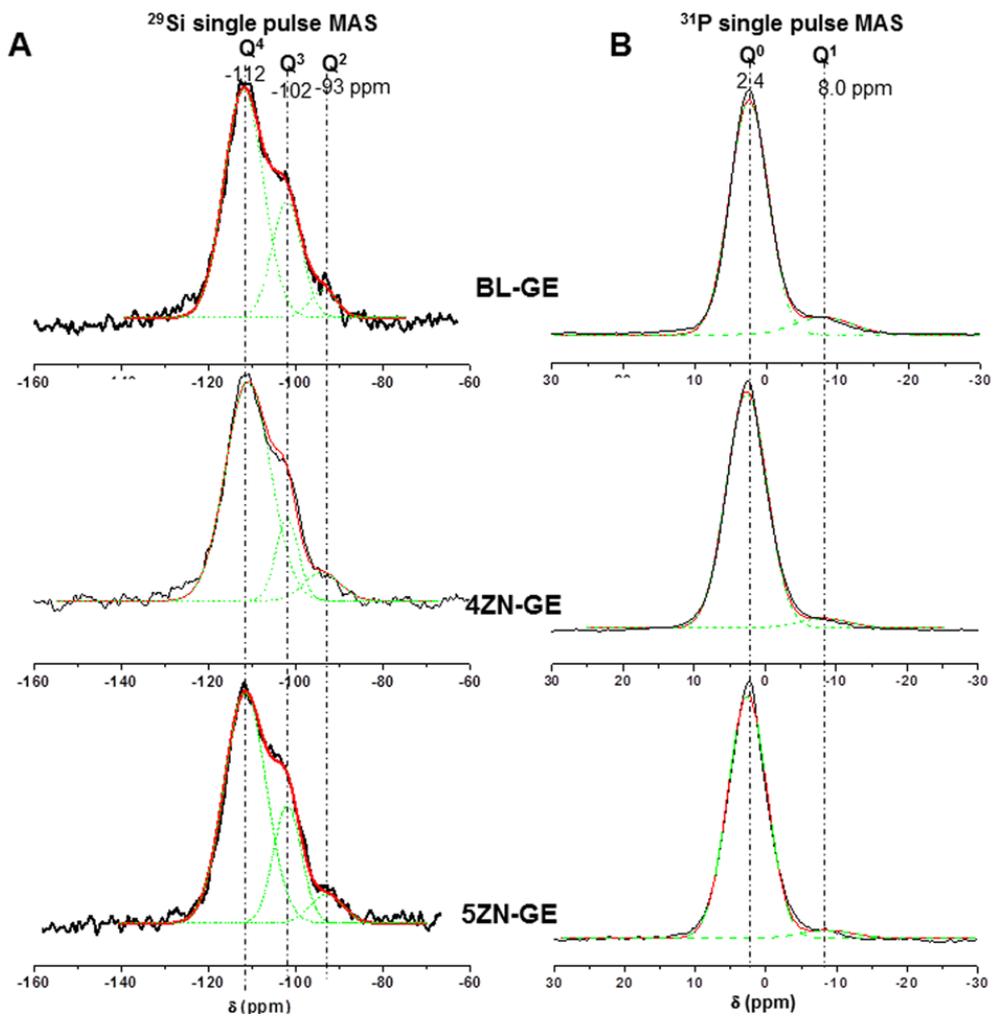

Figure 6

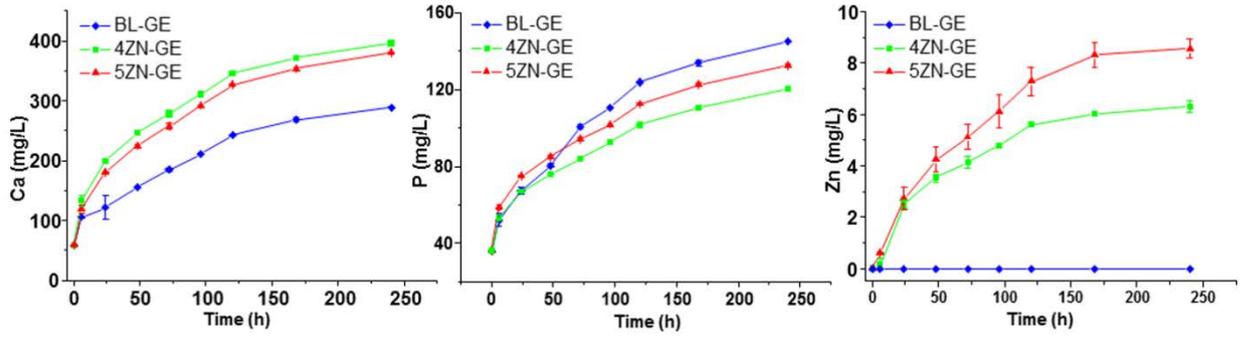

Figure 7

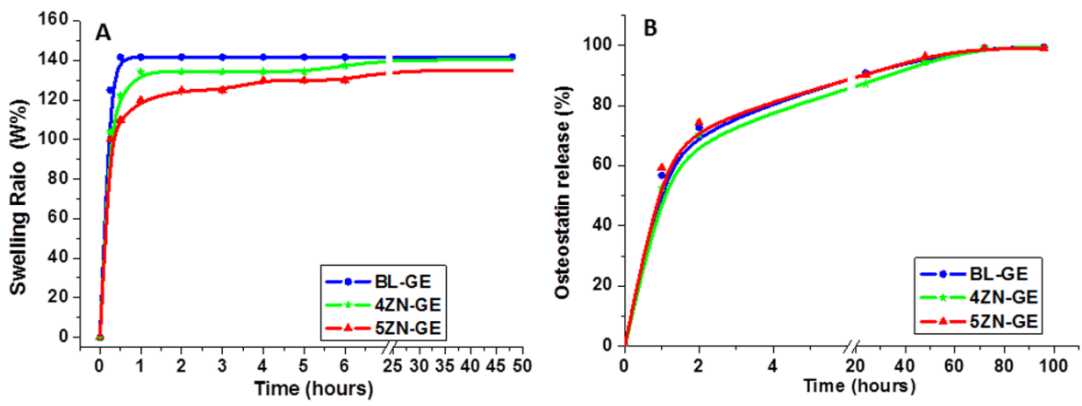

Figure 8

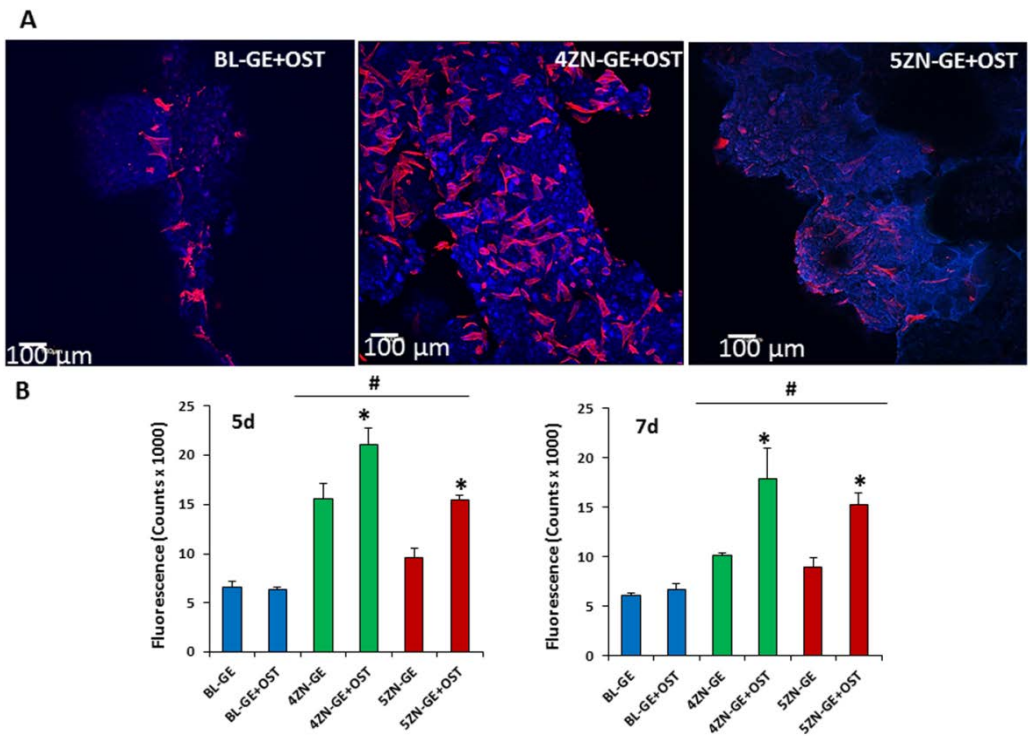

Figure 9

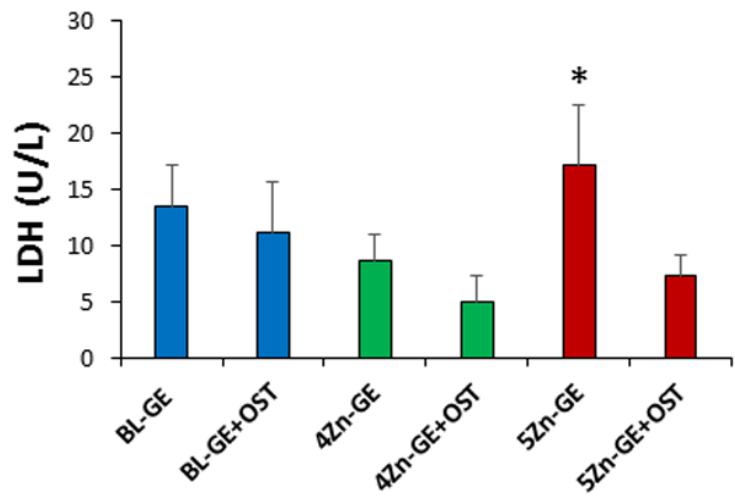

Figure 10

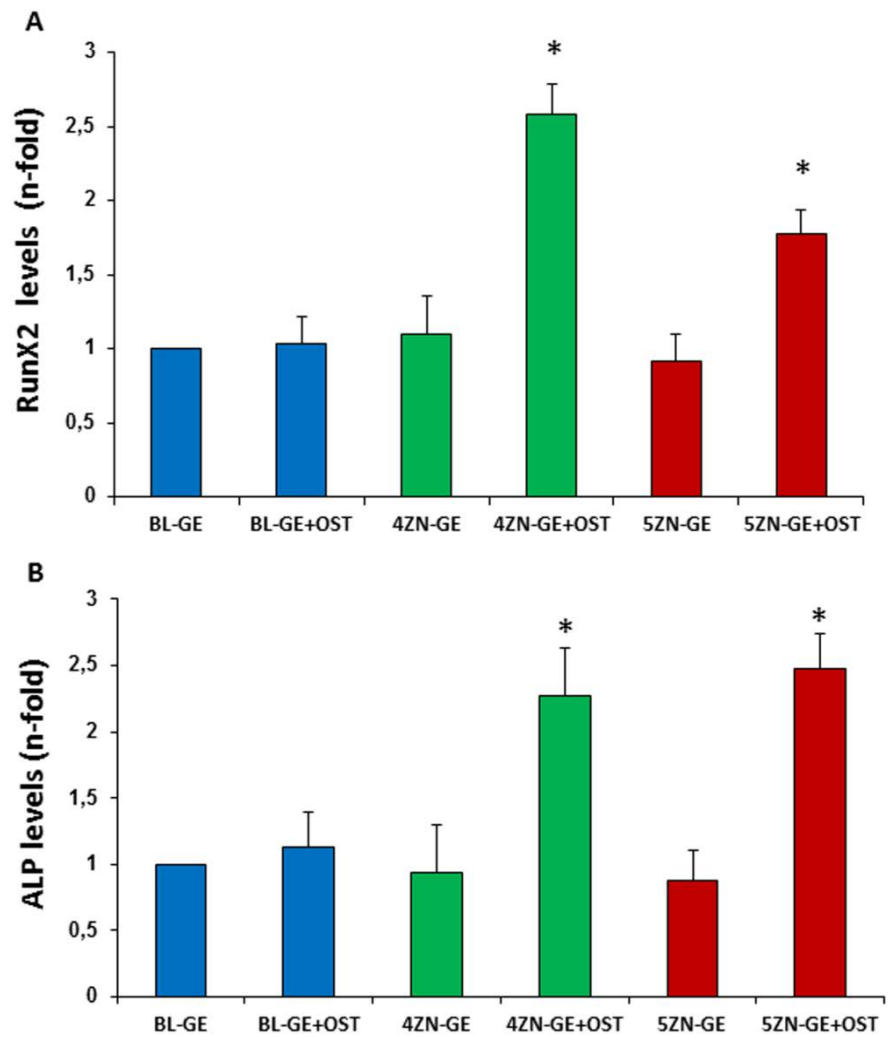

Figure 11

**Table 1**. Amounts of reactants used for the MBGs syntheses via EISA method.

| Sample | Pluronic® 123 (g) | Ethanol (mL) | HNO$_3$ (g) | TEOS (mL) | TEP (mL) | CaNO$_3$.4H$_2$O (g) | ZnNO$_3$.6H$_2$O (g) |
|---|---|---|---|---|---|---|---|
| **BL** | 4.5 | 67.5 | 1.12 | 8.90 | 0.71 | 1.10 | --- |
| **4ZN** | 4.5 | 67.5 | 1.12 | 8.90 | 0.71 | 1.10 | 0.60 |
| **5ZN** | 4.5 | 67.5 | 1.12 | 8.90 | 0.71 | 1.10 | 0.75 |

**Table 2.** Textural properties of the scaffolds being S$_{BET}$ the BET surface, V$_P$ pore volume and D$_P$ pore diameter. The percentage of PCL and gelatin contents in the scaffolds were determined by TGA. The weight percentages of Si, P, Ca and Zn in the scaffolds were determined by EDX. The nominal values are included in brackets.

| Sample | S$_{BET}$ (m$^2$/g) | V$_P$ (cm$^3$/g) | D$_P$ (nm) | %PCL (TGA) | % Gelatin (TGA) | Atomic Composition % (EDX) | | | |
|---|---|---|---|---|---|---|---|---|---|
| | | | | | | Si | P | Ca | Zn |
| **BL** | 342.9 | 0.40 | 5.1 | 0.5 (0) | -- | 84.5 (80) | 3.1 (5) | 12.5 (15) | -- |
| **BL-GE** | 188 | 0.21 | 3.7 | -- | 7 | -- | -- | -- | -- |
| **BL-OST** | 164 | 0.17 | 3.5 | -- | -- | -- | -- | -- | -- |
| **4ZN** | 238.1 | 0.40 | 3.3 | 1 (0) | -- | 82.2 (77) | 3.3 (4.8) | 10.3 (14.4) | 4.2 (4) |
| **4ZN-GE** | 151.4 | 0.14 | 3.5 | -- | 6 | -- | -- | -- | -- |
| **4ZN-OST** | 113.1 | 0.10 | 3.5 | -- | -- | -- | -- | -- | -- |
| **5ZN** | 188.4 | 0.14 | 3.3 | 1 (0) | -- | 82.5 (76) | 3.8 (4.8) | 8.9 (14.3) | 4.8 (5) |
| **5ZN-GE** | 135.5 | 0.14 | 3.5 | -- | 6 | -- | -- | -- | -- |
| **5ZN-OST** | 101.3 | 0.10 | 3.5 | -- | -- | -- | -- | -- | -- |

**Table 3.** Chemical shifts and relative peak areas obtained by solid state $^{29}$Si and $^{31}$P Single Pulse MAS NMR spectroscopy. The peaks areas for the $Q^n$ units were calculated by Gaussian line-shape deconvolutions and their relative populations were expressed as percentages (%).

| | $^{29}$Si | | | | | | | | | $<Q^n>^a$ | $^{31}$P | | | | | |
| --- | --- | --- | --- | --- | --- | --- | --- | --- | --- | --- | --- | --- | --- | --- | --- | --- |
| Sample | $Q^4$ | | | $Q^3$ | | | $Q^2$ | | | | $Q^0$ | | | $Q^1$ | | |
| | CS ppm | Area (%) | fwhm ppm | CS ppm | Area (%) | fwhm ppm | CS ppm | Area (%) | fwhm ppm | | CS ppm | Area (%) | fwhm ppm | CS ppm | Area (%) | fwhm ppm |
| BL-GE | -112 | 63.3 | 8.9 | -102 | 33.8 | 6.8 | -92 | 3.0 | 6.0 | 3.60 | 2.4 | 90 | 5.3 | 8.0 | 10.0 | 8.0 |
| 4ZN-GE | -111 | 67.4 | 10 | -102 | 28.7 | 5.0 | -92 | 3.8 | 8.3 | 3.63 | 2.6 | 95 | 5.7 | 8.2 | 5.0 | 7.3 |
| 5ZN-GE | -111 | 69.4 | 9.6 | -102 | 23.7 | 6.4 | -93 | 6.9 | 7.6 | 3.62 | 2.6 | 96 | 5.5 | 8.7 | 4.0 | 7.2 |

$^a$ Network connectivity of MBGs as a function of chemical composition $<Q^n> = (4\times\%Q^4)/100 + (3\times\%Q^3)/100 + (2\times\%Q^2)/100 + (\%Q^1)/100$

**Table 4.** Kinetic parameters of the osteostatin release from BL-GE, 4ZN-GE and 5ZN-GE scaffolds. $w_0$: initial mass loaded (μg OST/g MBG); $(w_t/w_0)_{max}$: maximum amount of the relative release of OST; $k_1$: release rate constant; δ: kinetic non-ideality factor; R: goodness of fit.

| Sample | $w_0$ (μg/g) | $(w_t/w_0)_{max}$ (%) | $k_1(\times10^3)$ (h) | δ | R |
| --- | --- | --- | --- | --- | --- |
| BL-GE | 56 | 96.2 ± 3.4 | 40.9 ± 2 | 0.48 ± 0.02 | 0.996 |
| 4ZN-GE | 52 | 98.2 ± 3.1 | 37.2 ± 1 | 0.50 ± 0.03 | 0.998 |
| 5ZN-GE | 59 | 98.9 ± 3.0 | 40.9 ± 4 | 0.47 ± 0.04 | 0.996 |